
\magnification=1200
\vskip .25in
\hskip 1.0in

\magnification=1200
\vskip .25in
\hskip 1.0in%

{\vbox{\centerline{\bf An Extended N=1 Fermionic Supercurrent and}
\centerline{\bf its Associated N=2 Superconformal Algebra  }}

\smallskip
\centerline{David M. Pierce}
\bigskip\centerline{\it Institute of Field Physics}
\centerline{\it Department of Physics and Astronomy}
\centerline{\it University of North Carolina}
\centerline{\it Chapel Hill, NC 27599-3255, USA}
\bigskip
An extended free fermionic construction of the internal N=1 world sheet
supercurrent for four-dimensional superstring theory is given. We show how it
can describe theories with massless fermions, and we discuss the corresponding
  N=2  superconformal alg
ebra. As an intermediate step, we show that an internal N=2 global
superconformal invariance occurs in {\it any} superstring theory with massless
fermions at tree level. To demonstrate this fact, we give the N=2 supercurrents
for a model with N=1 space-ti
me supersymmetry and a model without space-time supersymmetry.

\vskip20pt

\centerline{\bf 1. Introduction}
\vskip10pt
The internal conformal system of a superstring theory, and in particular the
form of the world sheet superVirasoro generator, determine important properties
of the low energy spectrum, including the presence of massless fermions and the
Yang-Mills gauge g
roup. In this letter, we give a new extended expression for the internal N=1
supercurrent for free fermions. It includes a background charge term as well as
an additional term proportional to the world sheet fermions; and it is more
general than both the
standard supersymmetric Feigin-Fuchs construction[1] and the general form
constructed out of 3n bosons[2]. Of course, two-dimensional field theories can
be described in terms of either fermionic or bosonic fields whenever the zero
modes of the bosonic fie
lds have the values $\sqrt{2\alpha'}p\in  Z+\nu$ where Z is the set of integers
and $\nu$ is a rational number. In sect. 2, we derive the supercurrent, exhibit
its hermitian form, and discuss its relevance to Yang-Mills gauge symmetries.
In order to furth
er analyze the role of this current in superstring models, we consider in sect.
3 the connection between world sheet properties and space-time properties. We
show that {\it any} superstring theory with massless fermions at tree level
has an N=2 internal
superconformal algebra. This N=2 superconformal algebra corresponds to a global
world sheet symmetry(not gauged) and so there is no corresponding ghost system.
Although it is well known[3,4] that space-time supersymmetry requires an N=2
world sheet supers
ymmetry, it is perhaps not so widely appreciated that string theories can have
N=2 world sheet supersymmetry but not tree level space-time supersymmetry. To
exemplify this feature, we give explicit constructions of the N=2 supercurrents
for a model with s
pace-time supersymmetry and for a model without space-time supersymmetry in
sect's. 4 and 5. In sect. 6, we discuss how the new fermionic construction of
the N=1 supercurrent can be used in  theories with massless fermions and we
compute its associated N=
2 supercurrents.

\vskip20pt
\centerline{\bf  2. An Extended Realization of the N=1 superconformal algebra}
\vskip10pt

The form of the N=1 supercurrent is critical in determining the particle
spectrum of the string theory. Therefore, new constructions of the supercurrent
may lead to more realistic string models.
The operators in the internal superconformal field theory that we would like to
consider are the world sheet fermions, $\psi^a(z)$ and the spin fields,
$\Sigma^A(z)$. We must construct the  supercurrent from various combinations of
these fields and their
derivatives so that the conformal weight is 3/2.
A hint of what these operators might be is given by the operator product of
spin fields
$\Sigma^A(z)\Sigma^B(w)=(z-w)^{-3/4}\tilde\psi^{AB}+(z-w)^{1/4}({\textstyle{1\over 3 !}}\tilde\psi\tilde\psi\tilde\psi+{\textstyle{1\over 2}}\partial\tilde\psi)^{AB}+.
..$
where $\tilde\psi\equiv \sqrt{1/2}\Gamma_a^{AB}\psi^a(w)$ and products of
$\tilde \psi$ are antisymmetric products of gamma matrices. These operator
products contain similar combinations of operators to those in the supercurrent
shown in (2.1).
 The supercurrent is:

$$T_F(z)=-{i\over 6}f_{abc}\psi_a(z)\psi_b(z)\psi_c(z)
+\rho_a\partial_z\psi_a(z) +\sigma_a\psi_a(z)z^{-1}\eqno(2.1a)$$
The corresponding Virasoro current is:
$$L(z)={\textstyle{ 1\over
2}}\partial_z\psi_a(z)\psi_a(z)+\rho_a\partial_zT_a(z)+
2\sigma_aT_a(z)z^{-1}+2(\sigma^2-\sigma_a\rho_a)z^{-2}\eqno(2.1b)$$
where $T_a=-{i\over 2}f_{abc}\psi_b\psi_c$,
and $f_{abc}$ are the structure constants of a Lie group,
$f_{abc}f_{abe}=2\delta_{ce}$, and $\rho$ and $\sigma$ are arbitrary vectors
subject to the restriction that $f_{abc}\rho_b\sigma_c=0$; and we find that
$c={{\rm dim}g\over 2}-12\rho^2$. These cons
traints are required for closure of the N=1 algebra.

In physical contexts, one usually wants $L_n^{\dagger}=L_{-n}$ and
$G_r^{\dagger}=G_{-r}$. This restriction is unnecessary mathematically but it
can be met by the following constraints. With the supercurrent in the form
$T_F(z)={1\over 2}\sum_r G_r z^{-r-
{3\over 2}}$, the hermiticity relation $G_r^{\dagger}=G_{-r}$ requires that
$\rho$ is pure imaginary and that the imaginary part of $\sigma$ is one half
$\rho$. These constraints can be written as ${\rm Re}\rho_a=0$ and ${\rm
Im}\rho_a=2{\rm Im}\sigma_a$.

Combining the constraints for closure of the algebra with the constraints for
hermiticity, we find the following condition must hold: $f_{abc}( {\rm
Re}\sigma_b)({\rm Im}\sigma_c)=0$. The central charge is then given by
$c={dimg\over 2}+12({\rm Im}\rho)^2
$.

 The second term in (2.1a) provides a supersymmetric version of the
Feigin-Fuchs free field construction of the Virasoro algebra [1].
The currents  in (2.1) form an N=1 superconformal algebra and satisfy the
following operator product expansions:

$$T_F(z)T_F(w)={{c/6}\over (z-w)^3}+{{1\over 2}{L(w)}\over
(z-w)}\hskip10pt;\hskip10pt
L(z)T_F(w)={{3\over 2}T_F(w)\over (z-w)^2}+{\partial T_F(w)\over (z-w)}$$

$$ L(z)L(w)={{c/2}\over (z-w)^4} +{2L(w)\over (z-w)^2}+{\partial L(w)\over
(z-w)}\eqno(2.2)$$

We now address the interesting question of how these currents will influence
the gauge symmetry. The $\rho$ and $\sigma$ independent terms in (2.1a,b) have
the standard Kac-Todorov mixing with the set of gauge generators $T_a(z)$[5].
This describes an alg
ebra containing superKac-Moody generators with the appropriate conformal
dimension. The terms proportional to $\rho_a$ and $\sigma_c$ in (2.1) break the
symmetry to a subset of generators given by $\hat \rho_b T_b(z)+ {2\over z}\hat
\rho_b\sigma_a\delta_{
ba}$ where $f_{abc}\hat \rho_b\sigma_c=0$, $f_{abc}\hat \rho_b\rho_c=0$ and
$\delta_{ab}\hat\rho_a\rho_b=0$.

To describe the real world, any string theory must have massless fermions which
then acquire a mass by symmetry breaking. In the next section, we investigate
what constraints massless fermions put on the form this supercurrent.
\vskip20pt

\centerline{\bf 3.  N=2 superconformal algebra}
\vskip10pt
An important question in string theory is whether or not space-time
supersymmetry is required for a consistent finite theory. Efforts have been
made to develop string models without space-time supersymmetry[3,4]. It is
therefore useful to know what world
sheet properties are required for a realistic model that may not have
space-time supersymmetry. It has been shown that a global N=2 superconformal
symmetry, which is global in the sense that it does not have a corresponding
N=2 ghost system, is required f
or space-time supersymmetry[6,7]. What about string models without space-time
supersymmetry? We now point out from  previous arguments that an N=2 global
world sheet invariance follows directly from the presence of massless fermions
at tree level. It then
 follows that theories which have N=2 superconformal symmetry do not
necessarily  have space-time supersymmetry. It turns out that there are at
least three conditions that are required to guarantee space-time
supersymmetry[7,8,9].These include N=2 superco
nformal invariance, quantization of the U(1) charges, and a condition on the
spectral flow operator.

The N=2 superconformal algebra is given by:
$$ L(z)L(w)={{1\over 2}c\over (z-w)^4 }+{2L(w)\over (z-w)^2}+{\partial
L(w)\over z-w}+...$$
$$L(z)T^{\pm}_F(w)={{3\over 2}T^{\pm}_F(w)\over (z-w)^2}+{\partial
T^{\pm}_F\over (z-w)}+...\hskip10pt;\hskip10pt L(z)J(w)={J(w)\over (z-w)^2} +
{\partial J(w)\over z-w}+...$$
$$J(z)J(w)={{1\over 3}c\over (z-w)^2}+...\hskip10pt;\hskip10pt
J(z)T^{\pm}_F(w)={\pm T^{\pm}_F(w)\over z-w}+...$$
$$T^+_F(z)T^-_F(w)={{1\over 12}c\over (z-w)^3}+{{1\over 4}J(w)\over
(z-w)^2}+{{1\over 4}L(w)+{1\over 8}\partial J(w)\over z-w}+...$$
$$T^{\pm}_F(z)T^{\pm}_F(w)=0+...\eqno(3.1)$$
{}From (3.1), it can be seen that the N=1 superconformal algebra is a subset of
the N=2 superconformal algebra. This shows the relationship between the actual
N=1 superconformal gauge algebra and the N=2 superconformal algebra.

We now consider a free N=1 superconformal field theory with fermions.
The holomorphic part of the massless fermion vertex operator in the $-{1\over
2}$ picture is $V^{\alpha}_{-{1\over 2}}(z)= S^{\alpha}(z) e^{-{1\over
2}\phi(z)} \Sigma(z)$.
Here, $S^{\alpha}$ are spin fields of the Lorentz group SO(3,1) and $\Sigma$
is a spin field of conformal weight 3/8 of the internal conformal field
theory.
The spin fields create the ground states of the Ramond or twisted sectors from
the Neveu-Schwarz vacuum. In a free string theory, the spin field can be
bozonized and put in the general form(omitting cocycles),
$\Sigma(z)=:e^{i\nu_i\phi_i(z)}:$ where, for
fermionic descriptions, $\nu$ describes the boundary conditions of the world
sheet fermions. For this argument, since we are only considering one massless
fermion, the  cocycles are irrelevant. The operator product of an internal spin
field with its hermi
tian conjugate contains
a dimension one field J(z):
$$\Sigma(z)\Sigma^{\dagger}(w)=(z-w)^{-3/4}+(z-w)^{1/4}{\textstyle{1\over
2}}J(w)+...\eqno(3.2)$$ which satisfies:
$$J(z)\Sigma(w)={{\textstyle{3\over 2}}\Sigma(w)\over {z-w}}+... \hskip10pt
;\hskip10pt
J(z)\Sigma^{\dagger}(w)={{-\textstyle{3\over2}}\Sigma^{\dagger}(w)\over{z-w}}+...\eqno(3.3)$$
The derivation of (3.2) and (3.3)  goes as follows. With the spin field in the
bosonic form above, $\Sigma(z)\Sigma^{\dagger}(w)=(z-w)^{-{3\over
4}}[1+(z-w)(i\nu_i\partial \phi_i(w))+...]$. The U(1) current, which is
conventionally normalized as in (3.1),

is $J(z)=2i\nu_i\partial\phi_i(z)$. Then,
$J(z)\Sigma(w)=2\nu_i\nu_i{\Sigma(w)\over (z-w)}={{\textstyle {3\over
2}}\Sigma(w)\over (z-w)}$ since ${\nu_i\nu_i\over 2}=h=3/8$.
If the internal conformal field theory is non trivial, the space-time
supersymmetry algebra can be used to derive these operator products.

The U(1) current and spin field can be simplified by making a rotation of the
bosonic fields $\phi_i$ to a basis $H,H_1,...H_{i-1}$. We rotate the fields
$\phi_i$ so that the spin field and U(1) current are functions only of the
boson H. This requires tha
t $H={\textstyle{2\over \sqrt{3}}}\nu_i\phi_i$. So in the new basis, we have
$J(z)=i\sqrt{3}\partial H(z)$, $\Sigma=:e^{i{\sqrt{3}\over 2}H}:$, and
$\Sigma^{\dagger} =:e^{-i{\sqrt{3}\over 2}H}:$. (3.2) and (3.3) can also be
derived in this new basis.

 The N=1 supercurrent does not have a definite U(1) charge, but it can be split
into two parts :
$$\eqalign{T_F(z) &=e^{i\sqrt{1\over 3} H}\tilde T^+_F(z) +
e^{-i\sqrt{1\over 3} H} \tilde T^-_F(z)\cr & \equiv T^+_F(z)+
T^-_F(z)}\eqno(3.4)$$ which have charge $\pm 1$ under J:
$$J(z)T^{\pm}_F(z)={\pm T^{\pm}_F(w)\over z-w}+...\eqno(3.5)$$
Here, $\tilde T^+_F$ and $\tilde T^-_F$ are fields  of conformal dimension 4/3.
The fields, $T_F^+$ and $T_F^-$ , satisfy the N=2 superconformal algebra,
(3.1). The form of the supercurrent in (3.4) can be derived by expressing the
supercurrent in a general form given by
$T_F=\Sigma_q {\rm exp}[i\sqrt{1/3}Hq]\tilde T^q_F +\Sigma_{\hat q}\partial
H{\rm exp}[i\sqrt{1/3}H\hat q]\hat T_F^{\hat q}$.

BRST invariance of the
massless fermion vertex operator and dimensional analysis require that the
operator product between $T_F$ and the spin fields contain a branch cut of
order 1/2,
$$T_F(z)\Sigma(w)=(z-w)^{-1/2}\tilde \Sigma (w)+...\hskip10pt;\hskip10pt
T_F(z)\Sigma^\dagger (w)=(z-w)^{-1/2}\tilde \Sigma ^\dagger (w)+...\eqno(3.6)$$
where $\tilde \Sigma$ and $\tilde \Sigma^\dagger $ are operators of dimension
11/8. Inserting the expressions for the supercurrent, $T_F=\Sigma_q {\rm
exp}[i\sqrt{1/3}Hq]\tilde T^q_F+\Sigma_{\hat q}\partial H{\rm
exp}[i\sqrt{1/3}H\hat q]\hat T_F^{\hat q}$
, and  the spin field, $\Sigma=e^{i{\sqrt{3}\over 2}H}$, into (3.6),
we find that the allowed charges are $q=\pm 1$, and $\hat T_F^{\hat q}=0$.
This tells us that the supercurrent can  be split into parts having definite
U(1) charges as in (3.4). To show that this form leads to an N=2 superconformal
algebra , we must determine the algebra generated by these currents.

The argument proceeds as discussed in [6]. For the sake of brevity, we only
give a brief outline of the closure of the N=2 algebra.

It now follows that
$$J(z)T_F(w)={{T_F^+ - T_F^-}\over (z-w)} \equiv {T_F^{\prime}\over
(z-w)}+...\eqno(3.7)$$ and
$-2T_F'$ defined here is the upper component of a dimension one N=1
superfield whose lower component is J(z)[10]. It follows from the
operator product expansion (OPE) of the superfields that
$T_F(z)T_F'(z)=-(z-w)^{-2}{1\over 2}J(w)-(z-w)^{-1}{ 1\over 4}\partial
J(w)+...$.
The OPE of $T_F' T_F'$ can be derived using the OPE'S
of $JJ, JT_F, JT_F'$, and $T_FT_F'$. It is given by
$T_F^{\prime}(z)T_F^{\prime}(w)=-T_F(z)T_F(w)+...$.
Finally, using the OPE'S $T_F'T_F', T_FT_F'$, and $T_F T_F$, we get the
operator product expansion for $T_F^+T_F^-$ given in (3.1). This closes the N=2
superconformal algebra.
 Therefore, any free string theory with N=1 world-sheet invariance and massless
fermions,i.e., any theory with a singularity of at most $-{1\over 2}$ as in
(3.6),  requires N=2 superconformal invariance. Furthermore, we see that there
is a different representation of the superconformal algebra associated with
each massless fermion. N
ote that this argument did not depend on any particular form of the action and
is valid for any free conformal field theory with N=1 superconformal symmetry
and massless fermions. The above argument also shows that in the more general
case of a nontrivial
 space-time metric, space-time supersymmetry requires an N=2 superconformal
invariance[6,7].

The vertex operators of the theory are primary fields of the N=2 superconformal
algebra. The states of these theories with an extended N=2 superconformal
algebra will satisfy the following conditions:
$$G_r^{\pm}|\phi>=0\hskip20pt r>0\eqno(3.8a)$$
$$L_n|\phi>=0\hskip20pt n>0\eqno(3.8b)$$
$$L_0|\phi>=h|\phi>\eqno(3.8c)$$
$$J_n|\phi>=0\hskip20pt n>0\eqno(3.8d)$$
$$J_0|\phi>=q|\phi>\eqno(3.8e)$$
q is the U(1) charge of the state. The conditions 3.8 b,d follow from
$G^{\pm}_r|\phi>=0$ and the N=2 algebra. Note however, that the U(1) charges of
all the physical states in a theory with N=2 (gauged) world sheet supersymmetry
would be zero whereas the
ories with a ``global'' N=2 world sheet supersymmetry can have nonzero U(1)
charges as in (3.10e).

If the string model with local N=1 world sheet supersymmetry has space-time
supersymmetry, there are constraints on the U(1) charges of the states.
Locality of the theory requires that the operator product of any two vertex
operators has no branch cuts. T
he presence of the gravitino in the space-time supersymmetric case requires
that the U(1) charges of all the states be integral or half-integral. In
general, the complete Hilbert space, containing the left and right movers, must
obey the locality conditio
n. In the case of the gravitino, the left movers always contribute integer
powers of $(\bar z -\bar w)$ so the right movers, which contain the U(1)
current J(z), must have integer powers of $(z-w)$. This is the reason that
there is a quantization conditio
n in the space-time supersymmetric case but not in a case without space-time
supersymmetry. To demonstrate this, we will give an example of a model which
has a conventional supercurrent and an N=2 superconformal symmetry, but does
not have  space-time sup
ersymmetry.

\vskip20pt

\centerline{\bf 4.  N=2 supercurrents for a chiral N=1 spacetime supersymmetric
model}
\vskip10pt

We now show explicit realizations  of the N=2 superconformal algebra for
particular type II superstring models. We give a general method for finding the
N=2
supercurrents provided certain criteria are met. Then, we relate the
expressions given in the fermionic formalism
below with the general forms given in the bosonic description in sect. 3.

We  consider four-dimensional type II models in the light-cone description.
The left and right movers can each be described by two bosonic and twenty
fermionic fields [11-14].
An N=1 model with $SU(2)\times U(1)^5$ gauge symmetry [12-15] can be described
by three generators $b_0,b_1,b_2$.The vectors,
$\rho_{b_i}=2(\nu_1,...\nu_n;\nu_1,...\nu_m;\nu_1,...\nu_{n'};\nu_1,...\nu_{m'})$ describing the boundary conditions of the fermi
ons in the respective sectors are:
\vskip1pt
\noindent$\rho_{b_0}=((1)^{12};(1)^4;(1)^{12};(1)^4)$,
$\rho_{b_1}=((0)^{12};(0)^4;(1)^4(0)^8;(0)^2(1)^2)$, and
$\rho_{b_2}=((0)^{10}(1)^2;$\vskip1pt
\noindent$(1/2)^4;(0)^2(1)^2(0)^4(1)^4;(1/2)^4)$.
The symbol h is used below to denote a real fermion that can have Ramond (R) or
Neveu-Schwarz (NS) boundary conditions.

The N=1 supercurrent  is given by:

$$\eqalign{T_F=&{i\over 2}h^1h^3h^7 +{i\over 2}h^2h^4h^8
+{1\over 2\sqrt 2}(h^5f^1\tilde f^3 + h^5f^3\tilde f^1 + h^6f^4\tilde f^2 +
h^6f^2\tilde f^4)\cr &
+{1\over 2\sqrt 2}(h^9\tilde f^1\tilde f^3 + h^9f^3f^1+h^{10}\tilde
f^2\tilde f^4 +h^{10}f^4f^2).}\eqno(4.1)$$

Using the N=2 superconformal algebra (3.1) we get $T_F^+$ and $T_F^-$ in terms
of $T_F$ and $T_F'$:
$$T_F^+={\textstyle{1\over 2}}(T_F + T_F');\qquad
T_F^-={\textstyle{1\over 2}}(T_F - T_F').\eqno(4.2)$$
Our basic approach to finding the N=2 supercurrents is to first determine a
U(1) current satisfying the N=2 superconformal algebra. This is determined from
either (3.2) or (3.3). The operator product of
 J(z) with the N=1 supercurrent will give $T_F'(w)$ from (3.7). Then, we use
(4.2) to put these N=2 supercurrents in the basis defined by $T_F^+$ and
$T_F^-$.

We now use the operator products of spin fields to determine a U(1) current of
the N=2 superconformal algebra. These operator products are most easily
calculated using the bosonization techniques introduced in [16].
The U(1) current associated with the spin field $\Sigma(z)=e^{{i\over
2}\phi_1}e^{{i\over 2}\phi_2}e^{{i\over 2}\phi_3}$ is $J(z)=i\partial
\phi_1(z)+i\partial \phi_2(z)+i\partial\phi_3(z)$. Since this
spin field is in the Ramond sector, the corresponding fermions will have Ramond
boundary conditions.
Now to relate this to the models with twisted fermions, we fermionize in the
following way:
$$e^{\pm i\phi_1}={1\over \sqrt 2}(h^1\mp i h^2).\eqno(4.3)$$
For the complex part, we bosonize the fermions in the following way:
$$e^{i\phi_2}= f^3 ;\hskip5pt e^{-i\phi_2}=\tilde f^3;\hskip5pt
e^{i\phi_3}=f^4;\hskip5pt
e^{-i\phi_3}=\tilde f^4.\eqno(4.4)$$
The U(1) current in fermionic form is then $$J(z)=i:h^1(z)h^2(z):+:
f^3(z)\tilde f^3(z):+: f^4(z)\tilde f^4(z):-{\nu_3\over z}-{\nu_4\over
z}.\eqno(4.5)$$

The N=2 supercurrents are then found using  (4.2).
$$T_F^+={i\over 4}h^1h^3h^7+{i\over 4}h^2h^4h^8+{1\over 4}h^2h^3h^7-{1\over
4}h^1h^4h^8
+{1\over 2\sqrt 2}\{f^3(\tilde f^1h^5+f^1h^9)+f^4(\tilde
f^2h^6+f^2h^{10})\}\eqno(4.6)$$
and $T_F^-={T_F^+}^{\dagger}$.

These results are consistent with the general result given in sect. 3 . There,
it was shown that whenever there is N=1 world-sheet supersymmetry and massless
fermions are present, the supercurrent can be put in the form of
$T_F=\sum_q{\rm exp}[i\sqrt{1/3}
H q]\tilde T^q_F$  with charges $\pm 1$. We  relate this general form to the
explicit form constructed from fermionic models.
We first bosonize the fermions of the U(1) current as in (4.3). The
supercurrent will then be
in a form given by
$T_F={\Sigma_{q_i}}e^{iq_1\phi_1}e^{iq_2\phi_2}e^{iq_3\phi_3}\tilde
T_F^{(q_i)}$
where $T_F$ has conformal dimension one and the allowed values of
$(q_1,q_2,q_3)$ are given by $(\pm 1,0,0)$,$(0,\pm 1,0)$, and $(0,0,\pm 1)$.

With the following rotation of the bosonic fields $\phi_1$,$\phi_2$,$\phi_3$

$$\phi_1={1\over \sqrt 3}(H+aH_2+bH_3);\phi_2={1\over \sqrt
3}(H+cH_2+dH_3);\phi_3={1\over \sqrt 3}(H+eH_2+fH_3)\eqno(4.7)$$

the supercurrent takes the form:

$$T_F=e^{i{\sqrt{1\over 3}}H}\tilde T_F^+ + e^{-i{\sqrt{1\over 3}}H}\tilde
T_F^-$$

$$\tilde T_F^+={e^{i(aH_2+bH_3)}\over 2\sqrt 2}(-h^4h^8+ih^3h^7)
+{e^{i(cH_2+dH_3)}\over 2\sqrt 2}(\tilde f^1h^5+f^1h^9)
+{e^{i(eH_2+fH_3)}\over 2\sqrt 2}(\tilde f^2h^6+f^2h^{10})\eqno(4.8)$$
and $\tilde T_F^- =\tilde T_F^{+\dagger}$.

The a,b,c,d,e, and f describe a rotation of the bosonic fields $\phi_i$
which put  the N=2 supercurrents and U(1) current in the desired form. The
U(1)
current is then given by $J(z)=i\sqrt{3}\partial H =i\partial \phi_1 +i\partial
\phi_2 +i
\partial \phi_3$ and the spin field is  $\Sigma=e^{i{\sqrt{3\over 2}} H}$.
These are the expressions given in sect. 3.

\vskip20pt

\centerline{\bf 5. N=2 supercurrents  for a string
 model without space-time supersymmetry}
\vskip10pt
In sect. 3, we outlined an argument that any string theory with N=1 world sheet
supersymmetry and massless fermions at tree level has N=2 superconformal
invariance.  In this section, we show an explicit construction of  the N=2
supercurrents for a superst
ring model with massless fermions but without space-time supersymmetry.  The
model we would like to consider is given in ref.[12.13] and is generated by the
sectors $b_0$,$b_1$ with
$\rho_{b_0}=((1)^{14};(1)^3;(1)^{14};(1)^3)$ and
$\rho_{b_1}=((0)^{12},(1)^2;({1\over 3})^3;(1)^2,(0)^8,(1)^4;{1\over
3},(-{2\over 3})^2)$. The right moving fields can be denoted by $h^{\hat
i}(z)$, $h^I(z)$,$f(z)=\sum_{r\in Z+\lambda}f_r z^{-r-{1\over 2}
}$, and $f^i(z)=\sum_{r\in z+\lambda}f_r^i z^{-r-{1\over 2}}$, where $\hat i,i$
=1,2; and I=1,...12. The associated supercurrent has c=9 for the internal
degrees of freedom and is given by
$$\eqalign{T_F=&-{i\over 2}[h^9h^1h^2+h^{10}h^3h^4+h^{11}h^5h^6]+{1\over 2
\sqrt{3}}[h^{12}(\bar h^7\bar h^8+\tilde f^2f^2+f^1\tilde f^1)\cr & +f(\bar
h^7\tilde f^2+\tilde f^1\bar h^8+f^1f^2)+\tilde f(\bar h^7f^1+f^2\bar
h^8+\tilde f^2\tilde f^1)]}\eqno(5.1)$$
where $\bar h^7=h^7+ih^8$, $\bar h^8=h^7-ih^8$.

The U(1) current derived from the internal spin field in sector $b_1^3$ has a
simple form. The boundary conditions of the fermions in this sector are given
by: $\rho_{b_1^3}=((0)^{12},(1)^2;(1)^3;$\vskip1pt
\noindent $(1)^2,(0)^8,(1)^4;(1),(0)^2)$. In this sector, there are no twisted
world-sheet fermions.  The U(1) current associated with the massless fermion in
sector $b_1^3$ is given by $J(z)=i:h^9h^{10}:+i:h^{11}h^{12}:+
:f\tilde f: -{\nu z^{-1}}$. The N=2 supercurrents are $T_F^-={T_F^+}^{\dagger}$
where

$$\eqalign{T_F^+=&-{1\over 4}(ih^9+h^{10})(h^1h^2+ih^3h^4)+-{1\over
4}(ih^{11}+h^{12})[h^5h^6-{1\over \sqrt{3}}(\bar h^7\bar h^8+\tilde
f^2f^2+f^1\tilde f^1)]\cr & + {1\over 2\sqrt{3}}f(\bar h^7\tilde f^2+\tilde
f^1\bar h^8+f^1f^2)}\eqno(5.2)$$

This model  does not meet all of the conditions that are required to have
space-time supersymmetry. Although it has an N=2 superconformal symmetry, all
the states do not have integral or half integral U(1) charges. As pointed out
in sect. 3, it has N=2 su
perconformal symmetry simply because it has massless fermions.

 There
is, in fact,  a different representation of the N=2 superconformal algebra
associated with every massless fermion. In other words, each internal spin
field of the massless fermion vertex operator gives rise to a U(1) current
belonging to a different repre
sentation of the N=2 superconformal algebra. For example,
the generator sector $\rho_{b_1}$ has an internal spin field in the twisted
sector. The U(1) current derived from  this spin field is
$J(z)=i:h^9h^{10}:+i:h^{11}h^{12}:+{1\over 3}:f\tilde f:-{\nu
z^{-1}}+{2\over3}:f^1\tilde f^1:-{\nu_1 z^{-1}}+{2\over 3}:
f^2\tilde f^2:-{\nu_2 z^{-1}}$. This U(1) current has corresponding N=2
supercurrents which can be derived as shown above.

\vskip20pt
\centerline{\bf 6. Extended N=1 supercurrent in theories with massless
fermions}
\vskip10pt
We now consider the new form of the N=1 fermionic supercurrent (2.1) and
determine when it can be used in theories with massless fermions. In sect. 3,
we showed that any supersting theory with massless fermions at tree level must
form an N=2 superconforma
l algebra. To be more precise, the internal supercurrent for the movers with
the massless fermions must be able to be split into two parts with U(1) charges
$\pm 1$. To form the supercurrent for the movers with  massless fermions,
$\rho_a$ in (2.1a) must
be zero. Furthermore, the gauge symmetry for the movers with the massless
fermions must be abelian, i.e. the first term in (2.1) must have, for eg., for
real world sheet fermions, SU(2) structure constants, and the fermions must
have mixed NS and R boundary conditions [17].

We now give a simple illustration of the N=2 supercurrents associated with the
new form of the N=1 supercurrent. In this example, we give an explicit form of
the supercurrent for the movers in a sector which contains a massless fermion.
The fermions with
space-time indices and six internal fermions will have Ramond boundary
conditions. The remaining 12 internal fermions will have Neveu-Schwarz boundary
conditions. In this case, the U(1) current associated with the internal spin
field is given by $J(z)=\sum_{n=1}^3 i:\psi_{(2n-1)}(z)\psi_{2n}(z):$. The
N=1 supercurrent splits into 
parts as $T_F=T_F^+ + T_F^- $ where $T_F^{\pm}$ have  U(1) charges $\pm 1$.
Thus, the N=2 supercurrents associated with the new fermionic form of the N=1
supercurrent (2.1) are:

$$\eqalign{T_F^+=& \sum_{n=1}^3[-{{\textstyle {i\over 4}}}\psi_{
(2n-1)}\psi_{(4n+3)}\psi_{(4n+4)}-{\textstyle {i\over
4}}\psi_{(2n)}\psi_{(4n+5)}\psi_{(4n+6)}+\cr &{\textstyle{1\over
4}}\psi_{(2n-1)}\psi_{(4n+5)}\psi_{(4n+6)}-{\textstyle{1\over 4}}\psi_{
(2n)}\psi_{(4n+3)}\psi_{(4n+4)}+\cr & \sigma^{(2n-1)}{\psi_{(2n-1)}\over
\textstyle {2z}}+\sigma^{2n}{\psi_{(2n)}\over \textstyle
{2z}}+i\sigma^{(2n)}{\psi_{(2n-1)}\over \textstyle
{2z}}-i\sigma^{(2n-1)}{\psi_{(2n)}\over \textstyle{2z}}]}\eqno(6.1)$$

\noindent and $T_F^-=T_F^{+\dagger}$.

Furthermore, in closed superstring theories, the entire hermitian form of the
supercurrent (2.1) can be used to describe the supercurrent for the movers
without a massless fermion. In such a case, massless fermions can appear on the
other side.

\vskip20pt
\centerline{\bf 7. Conclusions and comments}
\vskip10pt

 The form of the supercurrent plays a critical role in determining the particle
content of a string theory. This paper identifies a new  extended version of
the N=1 supercurrent.
In order to better understand its role, we analyzed the relationship between
space-time and world sheet properties. We addressed the question of which world
sheet properties are needed to have a theory with massless fermions at tree
level which is not nec
essarily space-time supersymmetric. We found that when a free conformal field
theory with a local N=1 superconformal invariance has massless fermions, a
global N=2 superconformal symmetry is required and there is no quantization
condition on the U(1) char
ges of the states.
To illustrate these general findings, we gave an example of a string model with
N=2 superconformal invariance which does not have space-time supersymmetry. We
then discussed how the new fermionic construction of the N=1 supercurrent can
be used in theorie
s with massless fermions and we computed its corresponding N=2 supercurrents.
 The extended N=1 currents in (2.1) are useful in describing theories with
restricted gauge symmetry. It will be interesting to incorporate this form of
the superVirasoro generators in models with generalized GSO projections.
\vskip20pt
\centerline{\bf Acknowledgements}
The author would like to thank Louise Dolan for many helpful discussions.

\noindent This work is supported in part by the US Department of Energy under
grant
DE-FG-05-85ER-40219/Task A.
\vskip10pt
\centerline{\bf References}
\item{1.} B.L. Feigin and D.B. Fuchs, Funct. Anal. Prilozhen {\bf 16} (1982)
47;
        C.B. Thorn, \vskip1pt  Nucl. Phys. {\bf B248} (1984) 551; M.
Bershadsky, V. Knizhnik, and A. Teitelman, \vskip1pt Phys. Lett. {\bf 151B}
(1985) 31
\item{2.} W.Lerche, B.E.W. Nilsson, and A. Schellekens, Nucl. Phys. {\bf
B294}(1987) 136
\item{3.} K. Dienes, Nucl. Phys. {\bf B429} (1994) 533
\item{4.} L.J. Dixon and J.A. Harvey, Nucl. Phys. {\bf B274} (1986) 93
\item{5.} V.G. Kac and I.T. Todorov, Comm. Math. Phys. {\bf 102} (1985) 337
\item{6.} T. Banks, L.J. Dixon, D. Friedan, E. Martinec, Nucl. Phys. {\bf B299}
(1988) 613
\item{7.} C. Hull and E. Witten, Phys. Lett. {\bf B160} (1985) 398
\item{8.} W. Lerche and N.P. Warner, Phys. Lett. {\bf B205} (1988) 471
\item{9.} R. Blumenhagen and A. Wi\ss kirchen, hep-th X 9503129.
\item{10.} C. Lovelace, Phys. Lett. {\bf 135B } (1984) 75;
     E. Fradkin and A. Tseytlin, \vskip1pt Phys. Lett. {\bf 158B} (1985) 316;
     A. Sen, Phys. Rev. Lett. {\bf 55} (1985) 1846;\vskip1pt
     C. Callan, D. Friedan, E.Martinec and M. Perry, Nucl. Phys. {\bf B262}
(1985) 593
\item{11.} R. Bluhm, L. Dolan, P. Goddard, Nucl. Phys. {\bf B289} (1987) 364
\item{12.} R. Bluhm, L. Dolan, P. Goddard, Nucl. Phys. {\bf B309} (1988) 330
\item{13.} I. Antoniadis, C. Bachas, C. Kounnas, Nucl. Phys. {\bf B289} (1987)
87
\item{14.} H. Kawai, D. Lewellen, J.A. Schwartz, H. Tye, Nucl. Phys.
{\bf B299} (1988) 431
\item{15.} L.J. Dixon, V.S. Kaplunovsky, C. Vafa, Nucl. Phys. {\bf B294} (1987)
43
\item{16.} V.A. Kostelecky, O. Lechtenfeld, W. Lerche, S. Samuel, S. Watamura,
\vskip1pt Nucl. Phys. {\bf B288} (1987) 173
\item{17.} I. Antoniadis and C. Bachas, Nucl. Phys. {\bf B298} (1988) 586

\end